\documentclass[9pt,twoside,a4paper]{article}
\usepackage{graphicx}
\usepackage{verbatim}
\usepackage{amssymb}
\usepackage{amsmath}
\usepackage[english]{babel}
\usepackage{color}
\usepackage[margin=2cm]{geometry}
\usepackage{authblk}

\begin{document}

\title{Position Dilution of Precision: a Bayesian point of view}
\author[1]{Alexandra Koulouri}
 \author[2]{Ville Rimpil\"ainen}
 \author[ ]{Nathan D. Smith}
 \small{
\affil[1]{Laboratory of Mathematics, Tampere University, P.O.\ Box
692, 33101 Tampere, Finland, alexandra.koulouri@tuni.fi}
\affil[2]{Department of Physics, University of Bath, BA2 7AY, Bath,
UK}}

\maketitle
\begin{abstract} The expected position error  in many cases is far
from feasible to be estimated experimentally using real satellite measurements
which makes the model-based position dilution of precision (PDOP)
crucial in positioning and navigation applications. In the following
sections we derive the relationship between PDOP and position error
and we explain that this relationship holds as long as the model for
the observation errors represents the true sources of errors.
\end{abstract}
\section{Observation model and position statistics}
For a ground receiver at location
$\mathrm{r}=(r_x,r_y,r_z)^\mathrm{T}$, the positioning error at time
$t$ between $\mathrm{r}$ and a reference location $\mathrm{r}_0$,
denoted by $\Delta \mathrm{r}=(\Delta r_x,\Delta r_y,\Delta
r_z)^\mathrm{T}$, can be expressed through the linear system
\cite{Langley1999,Marques2018}
\begin{equation}\label{eq:linearSystem}
b= A\Delta\mathrm{r}+\varepsilon,
\end{equation}
where $b\in\mathbb{R}^S$ is a vector with the differences between
the measured and modelled pseudo-range values,
 $A=[\mathrm{n}_1,\ldots,\mathrm{n}_S]^\mathrm{T}\in\mathbb{R}^{S\times 3}$ where $\mathrm{n}_s$ is a
 unit column
 vector pointing from the modelled (approximated) receiver position to the satellite
 and $S$ is the total number of visible
satellites at a time instance $t$.
 Error $\varepsilon\in\mathbb{R}^S$
represents the measurement noise plus model errors and ionospheric
effects.

Considering Gaussian observation noise, i.e.
$\varepsilon\sim\mathcal{N}(0,\Gamma_\varepsilon)$ based on the
Baye's theorem, the mean of the positioning error is
\begin{equation}
\label{eq:generalExpectedPositioningError}
\widehat{\Delta\mathrm{r}}=
(A^{\mathrm{T}}\Gamma_\varepsilon^{-1}A)^{-1}A^{\mathrm
T}\Gamma_\varepsilon^{-1} b, \end{equation} and the covariance is
\begin{equation}\label{eq:PostCovarianceWeights}
\Gamma=(A^\mathrm{T}\Gamma_\varepsilon^{-1} A)^{-1}\end{equation}
when $S\geq 3$ and
$\mathrm{rank}(A^{\mathrm{T}}\Gamma_\varepsilon^{-1}A)\geq 3$.

\section{Position dilution of precision}
The position dilution of precision is a measure of the uncertainty
of the estimates $\widehat{\Delta \mathrm{r}}$ and it is
parameterized according to the statistical characterization of the
errors $\varepsilon$ and the linear model used for the position
estimates (eq.~\ref{eq:linearSystem}) \cite{MILBERT2008,Misra2010}.
In particular,  by denoting $\Gamma_{xx}$, $\Gamma_{yy}$ and
$\Gamma_{zz}$ the diagonal elements of the covariance
$\Gamma=(A^\mathrm{T}\Gamma_\varepsilon^{-1} A)^{-1}$
(eq.~\ref{eq:PostCovarianceWeights}), the standard deviations of
$\widehat{\Delta \mathrm{r}}$
(eq.~\ref{eq:generalExpectedPositioningError}) in X,Y and Z
direction are $\sigma_{x}=\sqrt{\Gamma_{xx}}$,
$\sigma_{y}=\sqrt{\Gamma_{yy}}$ and $\sigma_{z}=\sqrt{\Gamma_{zz}}$
respectively. The dilution of precision is defined as the normalized
(with a common scaling factor $\kappa$) root mean square (RMS) of
the standard deviations given by
\begin{equation}
\mathrm{RMS}=\sqrt{\sigma_x^2+\sigma_y^2+\sigma_z^2}=\sqrt{\Gamma_{xx}+\Gamma_{yy}+\Gamma_{zz}}=\sqrt{\mathrm{tr}(\Gamma)}=
\sqrt{\mathrm{tr}((A^\mathrm{T}\Gamma_\varepsilon^{-1}
A)^{-1})},\end{equation} where $\mathrm{tr}()$ denotes the trace of
a matrix. The general form of the position dilution of precision is
\begin{equation}\label{eq_generalPDOP}
\mathrm{PDOP}_{XYZ}=\frac{\mathrm{RMS}}{\sqrt{\kappa}} =
\frac{\sqrt{\mathrm{tr}((A^\mathrm{T}\Gamma_\varepsilon^{-1}
A)^{-1})}}{\sqrt{k}}= \sqrt{\mathrm{tr}((A^\mathrm{T}W A)^{-1})}
\end{equation}
where $W=\left(\frac{\Gamma_\varepsilon}{\kappa}\right)^{-1}$.
 Based on that ratio, we can see that the weights $W$ can be derived based on
 the prior knowledge about the error statistics.
Moreover, we can notice that as matrix $A$ and $W$ do not depend on
the measurements, but only on the geometry and the weighting scheme,
dilution of precision can be computed from the satellite orbital
information  without needing real measurements. 

In the following text, we  see that the scaling factor $\kappa$ is
defined based on the model of the observation error. More precisely,
for an observation error modeled with covariance $\Gamma_\varepsilon
= \gamma I^{S\times S}$,
\begin{equation}\mathrm{RMS}=\sqrt{\gamma\mathrm{tr}( (A^\mathrm{T}
A)^{-1})}.\end{equation} Therefore, the common scaling factor is
$\kappa={\gamma}$, $W=I^{S\times S}$ and
$\mathrm{PDOP}_{XYZ}=\sqrt{\mathrm{tr}(( A^\mathrm{T} A)^{-1)}}$.

Now, we can consider an observation error with covariance modelled
as
$$\Gamma_\varepsilon=\gamma I^{S\times
S}+\Gamma_{\varepsilon_\mathrm{kn}}$$ where
$\Gamma_{\varepsilon_\mathrm{kn}}$ is known (e.g. it has been
estimated using real measurements). Using the matrix inversion
lemma, we have that $\Gamma_\varepsilon^{-1}= \gamma^{-1}
(I^{S\times
S}-(\gamma
\Gamma_{\varepsilon_\mathrm{kn}}^{-1}+I^{S\times S})^{-1})$.
Inserting the previous expression in the RMS, we have that
\begin{equation}\mathrm{RMS}=\sqrt{\gamma\mathrm{tr}((A^\mathrm{T}U
A)^{-1})},\end{equation} where $U=I^{S\times
S}-(\gamma\Gamma_{\varepsilon_\mathrm{kn}}^{-1}+I^{S\times S})^{-1}$
and $\Gamma_\varepsilon^{-1}=\gamma^{-1} U$. Therefore, the scaling
factor $\kappa=\gamma$ and now
$W=
U$.  
%
Therefore, based on the previous two types of models for the error covariance, PDOP is given by
\begin{equation}\label{eq_generalPDOP_1}
\mathrm{PDOP}_{XYZ}=
\frac{\sqrt{\sigma_x^2+\sigma_y^2+\sigma_z^2}}{\sqrt{\gamma}}=\frac{\sqrt{\mathrm{tr}((A^{\mathrm{T}}\Gamma_\varepsilon^{-1}A)^{-1})}}{\sqrt{\gamma}}=
\frac{\sqrt{\mathrm{tr}(\Gamma)}}{\sqrt{\gamma}}.
\end{equation}

\section{Average position error and position dilution of
precision}\label{sec:PDOPandPositionError} Let us first define that
$\Delta\mathrm{r}=\mathrm{r}_{\mathrm{tr}}-\mathrm{r}_0$ is the
position difference between the true XYZ position vector,
$\mathrm{r}_{\mathrm{tr}}=(r_{x_{\mathrm{tr}}},r_{y_{\mathrm{tr}}},r_{z_{\mathrm{tr}}})$,
and the computed (approximate) one,
$\mathrm{r}_0=(r_{x_0},r_{y_0},r_{z_0})$, and
$\widehat{\Delta\mathrm{r}}=\hat{\mathrm{r}}-\mathrm{r}_0$ is the
position vector difference between the estimated position vector
$\hat{\mathrm{r}}=(\hat{r}_{x},\hat{r}_{y},\hat{r}_{z})$ and
$\mathrm{r}_0$.

 Based on the linear system (eq.~\ref{eq:linearSystem}) and the
observation error
$\varepsilon\sim\mathcal{N}(0,\Gamma_{\varepsilon})$, the estimated
position is given by
\begin{equation}\label{eq:positionEstimate_appendix}
\hat{\mathrm{r}}=\mathrm{r}_0+(A^{\mathrm{T}}\Gamma_\varepsilon^{-1}A)^{-1}A^{\mathrm{T}}\Gamma_\varepsilon^{-1}b.
\end{equation}

The statistical expectation, denoted by $\mathbb{E}[\cdot] $ and
covariance, denoted by $\mathrm{Cov}[\cdot]$, of $\hat{\mathrm{r}}$
are
\begin{align}
\mathbb{E}[\hat{\mathrm{r}}]
=&\mathrm{r}_0+(A^{\mathrm{T}}\Gamma_\varepsilon^{-1}A)^{-1}A^{\mathrm{T}}\Gamma_\varepsilon^{-1}\mathbb{E}[b]=\mathrm{r}_{\mathrm{tr}}
\\
\mathrm{Cov}[\hat{\mathrm{r}}]=&
(A^{\mathrm{T}}\Gamma_\varepsilon^{-1}A)^{-1}A^{\mathrm{T}}\Gamma_\varepsilon^{-1}\mathrm{Cov}[b]\Gamma_\varepsilon^{-1}A
(A^{\mathrm{T}}\Gamma_\varepsilon^{-1}A)^{-1}=\Gamma A^{\mathrm{T}}\Gamma_\varepsilon^{-1}\Gamma_\varepsilon^{\mathrm{true}}\Gamma_\varepsilon^{-1}A \Gamma,
\end{align}
where
$\mathbb{E}[b]=\mathbb{E}[A(\mathrm{r}_{\mathrm{tr}}-\mathrm{r}_0)+\varepsilon]$,
$\mathbb{E}[\varepsilon]=0$ and
$\mathrm{Cov}[b]=\Gamma_\varepsilon^{\mathrm{true}}$.

{ The position error vector between the true and the estimated
position vector is
$\mathrm{e}=(e_x,e_y,e_z)=\hat{\mathrm{r}}-\mathrm{r}_{\mathrm{tr}}$.
} {The statistical expectation  and covariance of the position error
are respectively
\begin{align}
\mathbb{E}[\mathrm{e}]=&\mathbb{E}[\hat{\mathrm{r}}]-\mathrm{r}_{\mathrm{tr}}=0\\
\mathrm{Cov}[\mathrm{e}]=&\mathrm{Cov}[\hat{\mathrm{r}}].
\end{align}
}

{ The average square magnitude of the position error, given by
$\|\mathrm{e}\|^2=e_x^2+e_y^2+e_z^2=(\hat{r}_x-r_{x_\mathrm{tr}})^2+(\hat{r}_y-r_{y_\mathrm{tr}})^2+(\hat{r}_z-r_{z_\mathrm{tr}})^2$,
is
\begin{equation}
\mathbb{E}[\|\mathrm{e}\|^2]=\mathbb{E}[e_x^2]+\mathbb{E}[e_y^2]+\mathbb{E}[e_z^2]=\mathrm{tr}(\mathrm{Cov}[\mathrm{e}])=\mathrm{tr}(\mathrm{Cov}[\hat{\mathrm{r}}])=\mathrm{tr}(\Gamma
A^\mathrm{T}\Gamma_\varepsilon^{-1}\Gamma^{\mathrm{true}}_\varepsilon
\Gamma_\varepsilon^{-1}A\Gamma).
\end{equation}

If the modelled error covariance $\Gamma_\varepsilon$ approximates
well the true observation error covariance
$\Gamma_\varepsilon^\mathrm{true}$, then
\begin{equation}
\mathbb{E}[\|\mathrm{e}\|^2] =
\mathrm{tr}((A^{\mathrm{T}}\Gamma_\varepsilon^{-1}A)^{-1})
=\mathrm{tr}(\Gamma).
\end{equation}
From (eq.~\ref{eq_generalPDOP_1}) we have that $\mathrm{PDOP}_{XYZ}
\propto \sqrt{\mathrm{tr}(\Gamma)}$ and therefore,
\begin{equation}
\mathbb{E}[\|\mathrm{e}\|^2]=\gamma\mathrm{PDOP}^2_{XYZ}
\end{equation}
where $\gamma$ is a scaling factor.}

Finally,
\begin{equation}\label{eq:PDOP_prop_error}
\mathrm{PDOP}_{XYZ}\propto
\sqrt{\mathbb{E}[\|\mathrm{e}\|^2]}.\end{equation}Therefore, we can
observe that the position dilution of precision is proportional to
the root of the expected square magnitude of the position error.
Equation (eq.~\ref{eq:PDOP_prop_error}) holds and thus PDOP can be
used to quantify the expected position errors under the condition
that
$\Gamma_\varepsilon\rightarrow\Gamma_\varepsilon^{\mathrm{true}}$.

In general, by using the (gross) covariance
$\Gamma_\varepsilon=\gamma I^{S\times S}$ when special ionospheric
weather take places, we end up with a PDOP which can give too
optimistic values because it neglects statistical information about
observation errors induced by the ionospheric conditions. For more
informative PDOP values, we should try to approximate
$\Gamma_\varepsilon\approx\Gamma_\varepsilon^{true}$. This
covariance can be estimated using statistical knowledge on
ionospheric phenomena (such as scintillation) extracted from real
measurements (e.g. $S_4$, $\sigma_\phi$ data as in
\cite{Koulouri2019})\footnote{We note that here we have assumed that
there is not bias or systematic errors and thus $
\mathbb{E}[\varepsilon]=0$. Similar analysis as in
section~\ref{sec:PDOPandPositionError} can be followed when   $
\mathbb{E}[\varepsilon]\neq 0$ resulting in a modified
(eq.~\ref{eq:PDOP_prop_error}) .
}.
\bibliographystyle{plain}
\bibliography{bib_scintillation}

\begin{thebibliography}{1}

\bibitem{Koulouri2019}
Alexandra Koulouri, Nathan~D. Smith, Bruno~C. Vani, Ville Rimpilainen, Ivan
  Astin, and Biagio Forte.
\newblock Methodology to estimate ionospheric scintillation risk maps and their
  contribution to position dilution of precision on the ground.

\bibitem{Langley1999}
Richard~B. Langley.
\newblock Dilution of precision.
\newblock {\em GPS World}, 10(5):52--59, 1999.

\bibitem{Marques2018}
Haroldo~Antonio Marques, Helo{\'{\i}}sa Alves~Silva Marques, Marcio Aquino,
  Sreeja~Vadakke Veettil, and Jo{\~{a}}o Francisco~Galera Monico.
\newblock Accuracy assessment of precise point positioning with
  multi-constellation {GNSS} data under ionospheric scintillation effects.
\newblock {\em Journal of Space Weather and Space Climate}, 8:A15, 2018.

\bibitem{MILBERT2008}
Dennis Milbert.
\newblock Dilution of precision revisited.
\newblock {\em Navigation}, 55(1):67--81, mar 2008.

\bibitem{Misra2010}
Pratap Misra and Per Enge.
\newblock {\em Global Positioning System: Signals, Measurements, and
  Performance (Revised Second Edition)}.
\newblock Ganga-Jamuna Press, 2010.

\end{thebibliography}

\end{document}